\documentclass[final,3p,times]{elsarticle}

\usepackage{graphicx}

\usepackage{amssymb}

\usepackage{amsmath}

\begin{document}

\begin{frontmatter}



\title{Energy spectrum of cosmic ray muons in $\sim$ 100 TeV energy region reconstructed from the BUST data}



\author[label1]{A.G. Bogdanov\corref{cor1}}
\ead{AGBogdanov@mephi.ru}
\author[label1]{R.P. Kokoulin}
\author[label2]{Yu.F. Novoseltsev}
\author[label2]{R.V. Novoseltseva}
\author[label2]{V.B. Petkov}
\author[label1]{A.A. Petrukhin}


\address[label1]{National Research Nuclear University MEPhI, 115409 Moscow, Russia}
\address[label2]{Baksan Neutrino Observatory INR RAS, 361609 Neutrino, KBR, Russia}
\cortext[cor1]{Corresponding author. Address: NEVOD, MEPhI, Kashirskoe sh. 31, Moscow 115409, Russia.}

\begin{abstract}
Differential and integral energy spectra of cosmic ray muons in the energy range from several TeV to $\sim$ 1 PeV obtained by means of the analysis of multiple interactions of muons (pair meter technique) in the Baksan underground scintillation telescope (BUST) are presented. The results are compared with preceding BUST data on muon energy spectrum based on electromagnetic cascade shower measurements and depth-intensity curve analysis, with calculations for different muon spectrum models, and also with data of other experiments.
\end{abstract}


\end{frontmatter}


\section*{Introduction}
\label{}

Energy spectrum of muons plays an important role in the physics of high energy cosmic rays. Its characteristics depend on the primary cosmic ray spectrum and composition, and also on the processes of primary particle interactions with nuclei of air atoms. Therefore information on muon energy spectrum may be used, on the one hand, for extraction of independent estimates of primary spectrum and composition if to suppose that interaction model is known, and, on the other hand, under certain assumptions about primary cosmic ray spectrum and composition, for the search of possible changes in characteristics of hadron interactions above the energy limit reached in accelerator experiments.

The region of muon energies above 100 TeV is of a special interest. In this region, the contribution of ``prompt'' muons from decays of charmed and other short-lived particles at reasonable suppositions about cross sections of their production can appear. This has to give some excess of such muons. But on the other hand the influence of the knee in the primary energy spectrum on muon spectrum shape is expected if the knee has really astrophysical origin. This effect leads to the decrease of muon flux at such energies. In the alternative case, if the spectrum of primary particles does not change its slope and appearance of the knee is connected with interaction model changes, the inclusion of new physical processes or states of matter is required. These processes can change the whole picture of EAS development, and standard estimations of EAS energies can appear wrong. As shown in \cite{1}, in this case the excess of muons will be increasing with energy rather sharply.

Usually, possible contribution of any fast (in comparison with decays of pions and kaons) processes to generation of muons is taken into account by means of introduction of the parameter $\textit{R}$ in the formula describing the inclusive spectrum of high-energy muons in the atmosphere \cite{2}:

\begin{equation}
\dfrac{dN_{\mu}}{dE_{\mu}}=0.14E_{\mu}^{-\gamma_{\mu}}\times\left( \dfrac{1}{1+\dfrac{1.1 E_{\mu}\cos \theta}{115}}+\dfrac{0.054}{1+\dfrac{1.1 E_{\mu}\cos \theta}{850}} + R\right)\:\:\:\mathrm{[cm^{-2}\:s^{-1}\:sr^{-1}\:GeV^{-1}].}
\end{equation}

Here $\textit{R}$ is the ratio of the number of these prompt muons to the number of charged pions with the same energy at production; muon energy is measured in GeV; $\gamma_{\mu} = 2.7$. The slope of the energy spectrum of prompt muons is about a unit less than that from decays of pions and kaons. In contrast to high energy muons from $\pi$-, $\textit{K}$-decays, the flux of prompt muons does not exhibit $\sec \theta$ enhancement with the increase of zenith angle. Unfortunately, cross sections of charmed particle production in a necessary range of kinematic variables are poorly known, and existing theoretical estimates of the  $\textit{R}$ value have a large spread. Nevertheless, the expected range of energies where the fluxes of prompt muons and muons from  $\pi$-, $\textit{K}$-decays become comparable is near 100 TeV \cite{3}; differential spectra of prompt muons and ``usual'' muons (from pion and kaon decays) are equal to each other at this energy for $\textit{R}\approx10^{-3}$.

If the observed knee in extensive air showers (EAS) energy spectrum at PeV energies is related with the inclusion of new physical processes (or formation of a new state of matter) with production in a final state of very high energy (VHE) muons \cite{4}, then their contribution to muon energy spectrum may be estimated by the formula:

\begin{equation}
\dfrac{dI_{\mathrm{VHE}\mu}}{dE_{\mu}}=\dfrac{A\gamma_{1}(E_{0}/10^{6}\:\mathrm{GeV})^{-\gamma_{1}}}{E_{0}}\times
\dfrac{n_{\mu}^{2}} {f_{\mu}\left[ 1-\left( \gamma_{1}/\gamma_{2}\right)\left( E_{k}/E_{0}\right)^{\left( \gamma_{2}-\gamma_{1}\right) /\gamma_{2}}  \right]},
\end{equation}
where $E_{\mu}$ and primary particle energy $E_{0}$ are related as
\begin{equation}
E_{\mu}\:n_{\mu}/f_{\mu}=\Delta(E_{0})=E_{0}-E_{k}\left( E_{0}/E_{k}\right)^{\gamma_{1}/\gamma_{2}},\:E_{0}>E_{k}.
\end{equation}

Here $\gamma_{1}$ and $\gamma_{2}$ are integral EAS energy spectrum slopes below and above the knee energy $E_{k}$; $\mathit{n_{\mu}}$ is a typical multiplicity of produced VHE muons; $\mathit{f_{\mu}}$ is a fraction of the difference between primary particle energy and measured EAS energy which is carried away by VHE muons; $A = 1.5\times10^{-10}\:\mathrm{cm^{-2}\:s^{-1}\:sr^{-1}}$ is the integral intensity of primary particles with energy above 1 PeV.

Appearance of VHE muons is also expected at energies about 100 TeV; however, their relative contribution should increase with energy more rapidly compared to muons from charmed particle decays, and this feature is the only one which could allow separate two hypotheses on possible reasons of changes in the muon energy spectrum behavior. There are few experimental data in the energy region close to 100 TeV (including the data obtained with BUST), and they have a very wide spread (see, e.g., review \cite{5}). This spread is most probably caused by various uncertainties of the methods used for investigations of the muon energy spectrum in the range $\geq$10 TeV.

Unfortunately, the most direct method of muon energy spectrum study -- magnetic spectrometer technique -- did not allow reach energies above 10 TeV because of both technical (the necessity to ensure high magnetic field induction simultaneously with manifold increase of magnetized volume) and physical (increase of probability of secondary electron contamination in the events with the increase of muon energy) reasons. Therefore, two other methods of muon spectrum investigations -- calorimeter measurements of muon-induced cascade shower spectrum and depth-intensity curve analysis -- were mainly used.

Method based on the depth-intensity measurements has serious uncertainties in estimation of the surface muon energy related with ambiguities in rock density and its composition, their non-uniformity in depth, and, in case of the mountain overburden, with errors in slant depth evaluation. Besides, this method has a principal upper limitation for accessible muon energies, since at depths more than about 12 km w.e. (in standard rock) the intensity of atmospheric muons becomes lower than the background flux of muons locally produced by neutrinos in the surrounding material. Taking into account energy loss fluctuations, such depth corresponds to effective muon threshold energy about 100 TeV.

The method based on calorimetric measurements of the spectrum of electromagnetic cascades induced via muon bremsstrahlung does not have upper physical limit. However, possibilities of investigations of muon spectrum at high energies are limited by low probability of the production of bremsstrahlung photons with energies comparable to muon energy ($\varepsilon_{\gamma} \sim E_{\mu}$), rapidly decreasing muon intensity, and consequently, by the necessity of corresponding increase of the detector mass. A special case of this technique represents the burst-size technique, when the cascade is detected in one point (in one layer of the detector). Such approach is used in the analysis of horizontal air showers (HAS) which may be produced deep in the atmosphere only by muons (or neutrinos). However, many questions appear in interpretation of measurements of this kind: Is the shower produced by single muon or by several particles? How to reject the background contribution from usual (hadron-induced) EAS? What is the effective target thickness for such observations? As a rule, there are no simple answers for these questions.

It is important to mark that, in contrast to magnetic spectrometer technique where the energies (the momenta) of individual particles are measured and differential energy spectrum may be directly constructed, two other methods provide essentially integral estimates: intensity of muons penetrating to the observation depth in depth-intensity measurements, and the amount of muons with energies exceeding the energy of bremsstrahlung photon in measurements of cascade shower spectrum. At that, effective muon energies do not strongly exceed the energy threshold (typically, about 2 times).

Since the methods discussed above encounter serious difficulties of principal or technical character, other methods are needed to ensure a breakthrough in the energy region $\sim$ 100 TeV and higher. From this point of view, the most promising seems to be pair meter technique \cite{6,7}. This method of muon energy evaluation is based on measurements of the number and energies of secondary cascades (with $\varepsilon<<E_{\mu}$) originated as a result of multiple successive interactions of muon in a thick layer of matter, mainly due to direct electron-positron pair production. At sufficiently high muon energies, in a wide range of relative energy transfers $\varepsilon/E_{\mu}\sim 10^{-1}-10^{-3}$ pair production becomes the dominating muon interaction process, and its cross section rapidly increases with $E_{\mu}$. A typical ratio of muon energy and the energy of these secondary cascades is determined by muon and electron mass ratio and is of the order of 100. An important advantage of this technique is the absence of principal upper limitation for measured muon energies (at least up to $10^{16}-10^{17}$ eV, where the influence of Landau-Pomeranchuk-Migdal effect on direct electron pair production cross section may become important). In case of a sufficient setup thickness ($\geq$ 500 radiation length) and large number of detecting layers (of the order of hundred) pair meter technique allows estimating individual muon energies; possibilities of the method for relatively thin targets depend on the shape of the investigated muon energy spectrum.

In the present paper, BUST data are analyzed on the basis of a modification of multiple interaction method elaborated for realization of the pair meter technique in thin setups. The results are compared with earlier BUST data on muon spectrum obtained by means of electromagnetic cascade shower measurements and depth-intensity curve analysis.

\section{Measurements of depth-intensity curve and spectrum of electromagnetic cascades at BUST}

BUST \cite{8} is located in an excavation under the slope of Mt. Andyrchy (North Caucasus) at effective rock depth $\mathrm{850\:hg/cm^{2}}$ which corresponds to about 220 GeV threshold energy of detected muons. The telescope (Fig. 1) represents a four-floor building with the height 11 m and the base $\mathrm{17\times17\:m^{2}}$. The floors and the walls are entirely covered with scintillation detectors (the total number 3152) which form 8 planes (4 vertical and 4 horizontal, two of the latter being internal ones). The upper horizontal plane contains $\mathrm{576=24\times24}$ scintillation detectors, the other three $\mathrm{400=20\times20}$ detectors each. The distance between neighboring planes in a vertical is 3.6 m. Total thickness of one layer (construction materials and scintillator) is about 7.2 radiation length.

\begin{figure}[h!]
\center \includegraphics[scale=0.5]{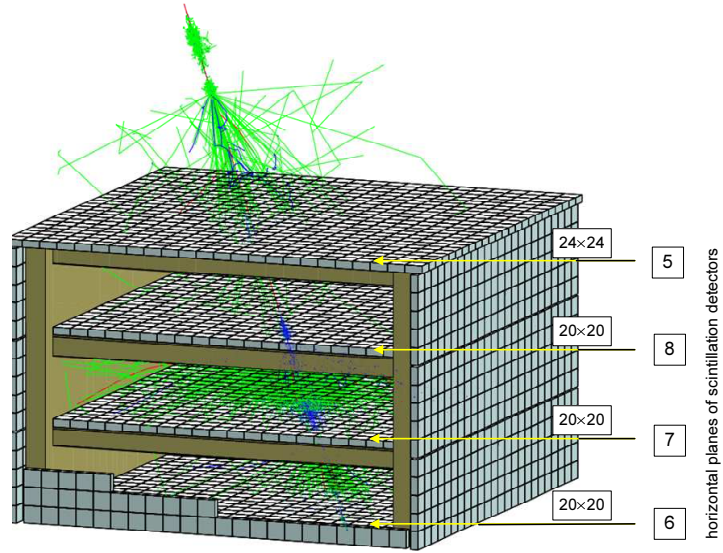}
\caption{High-energy muon passing through Baksan underground scintillation telescope (Geant4 simulation). Numbers of the planes (5 to 8) correspond to sequence of their construction.}
\end{figure} 

Each of the detectors represents an aluminum tank with sizes $\mathrm{0.7\times0.7\times0.3\:m^{3}}$ filled with liquid scintillator on the basis of white spirit viewed by a 15 cm diameter PMT (FEU-49) through PMMA illuminator. Most probable energy deposition in the detector at passage of a near-vertical muon is 50 MeV. The anode output of PMT serves for measurements of the energy deposition in the plane in the range from 12.5 MeV to 2.5 GeV and for the formation of master pulses for various physical programs. Pulse channel with operating threshold of 12.5 MeV (since 1991, 10 MeV threshold) is connected to $\mathrm{12^{th}}$ dynode and provides coordinate information (``yes-no'' type). The signal from $\mathrm{5^{th}}$ dynode of PMT is used to measure the energy deposition in the detector in the range from 0.5 to 600 GeV by means of the logarithmic converter of the pulse amplitude to duration.

BUST was created for investigations of cosmic ray muons and neutrinos as a telescope, but in principle it can detect as single muons (and muon bundles) so muon-induced cascade showers. Therefore, for the analysis of data concerning muon energy spectrum, three methods can be used: depth-intensity relation, measurements of the spectrum of electromagnetic cascades, and pair meter technique.

Results of the analysis of the BUST data on the depth-intensity dependence are given in \cite{9}. The underground muon intensity was measured in two zenith angle intervals ($50^{\circ}-70^{\circ}$ and $70^{\circ}-85^{\circ}$) for slant depth between 1000 and 12000 $\mathrm{hg/cm^{2}}$. Up to 6000 $\mathrm{hg/cm^{2}}$ in both zenith angle intervals the measured intensities agreed with the expectation for a usual muon spectrum (from pion and kaon decays). However at greater depths some excess of muons at moderate zenith angles ($50^{\circ}-70^{\circ}$) was observed which was interpreted by the authors as an indication for the appearance of prompt muons from charmed particle decays. The estimated contribution of prompt muons corresponded to the value of the parameter $R=(1.5\pm0.5)\times10^{-3}$.

Results of investigations of muon spectrum by means of measurements of the spectrum of electromagnetic cascades in BUST are described in \cite{10}. In this work, the muon energy spectrum $I_{H} (> E_{H})$ at the depth of setup location was derived from the spectrum of energy depositions in the telescope which was used as a 4-layer sampling calorimeter. For re-calculation to the surface muon energy spectrum $I_{0} (> E_{\mu})$ the authors used the solution of the kinetic equation for muon flux passing through a thick layer of matter. Some excess of the number of cascades in the tale of the spectrum found in this experiment could be caused as by methodical so by physical (inclusion of prompt muons) reasons. Authors noted that a similar flattening of the spectrum was also observed in a number of other experiments, but at different energies, which evidences in favor of methodical reasons of its appearance.

As a whole, results of the analysis of the BUST data on the depth-intensity curve and electromagnetic cascade spectrum poorly agree with each other. Below, results of independent analysis of the available BUST data based on ideas of the pair meter technique are described.

\section{Application of method of multiple interactions in BUST}

In order to evaluate individual muon energies (assuming that they have a usual power type integral spectrum $\sim E_{\mu}^{-2.7}$) by means of the pair meter technique with a reasonable accuracy, it is necessary to detect several ($\geq5$) muon interactions in the setup with total target thickness of several hundred radiation length and $\sim100$ detecting layers. If the number of layers and the setup thickness are low, the pair meter technique turns into the method of ``plural'' (in a limiting case, twofold) interactions. In this situation, evaluation of energies of individual muons is practically impossible; however, energy characteristics of the muon flux may be investigated on a statistical basis. The sensitivity of such method depends on the shape of muon energy spectrum and, as estimates show, for a more flat spectrum than the usual one, for example $N_{\mu}\sim E_{\mu}^{-1.7}$ (prompt muons, VHE muons, EAS muons in the range $E_{\mu}<<E_{0}$), it is sufficient to detect only two interactions even in the setup with the thickness of the order of several tens radiation length.

A significant volume of experimental data accumulated at BUST (more than 10 years of observations in combination with about $\mathrm{200\:m^{2}\:sr}$ geometric acceptance of the telescope) allows infer conclusions on the behavior of the muon spectrum in the region of very high energies on the basis of the method of multiple interactions, in spite of a small number of layers in the telescope (four) and low setup thickness ($\sim 30$ radiation length). 

In fact, the structure of BUST allows distinctively select not more than two successive interactions of muon in the telescope (Fig. 2). In the longitudinal profile of energy depositions (in horizontal planes) in such events, a minimum (``deep'', $E_{\textrm{min}}$) in one of the inner planes and two maximums (``humps'') above and below it must be observed. It is convenient to denote as $E_{1}$ the energy deposition measured in the higher maximum, $E_{2}$ the deposition in the second one; then the depth of the deep may be characterized by the ratio $K_{2}=E_{2}/E_\textrm{min}$.

\begin{figure}[h!]
\center \includegraphics[scale=0.6]{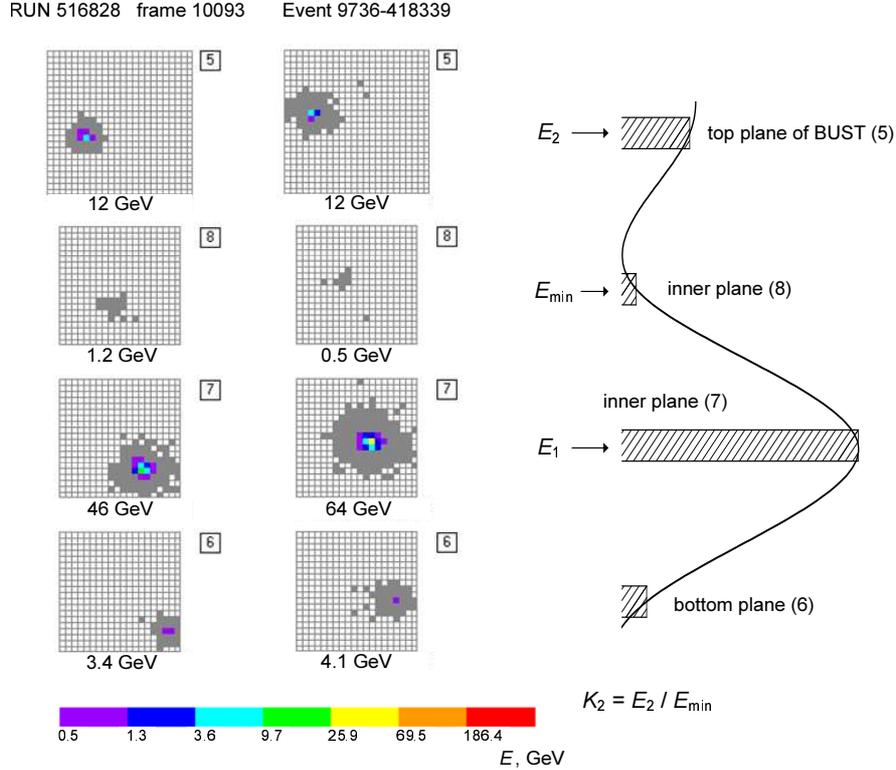}
\caption{Twofold interactions of muons in BUST. Left: examples of detected and simulated events; horizontal telescope planes are plotted. Hit detectors (``yes-no'') are shown in grey; colors correspond to different energy depositions (the scale below). Right: longitudinal profile of energy depositions in the telescope and definition of phenomenological parameters of the event.}
\end{figure} 

Simulation of the BUST response for the passage of single muons was performed by means of Geant4 toolkit \cite{11,12}. Before production of large-scale simulations, comprehensive tests of the correctness of muon electromagnetic interaction processes implementation in Geant4 in a wide range of energies and for various materials were done. The number of simulated events for muon energies above 350 GeV (at ground surface) was comparable to the expected number of such muons for the observation period (at ``usual'' energy spectrum), and for energies more than 1 TeV, 10 TeV, and 100 TeV exceeded the expected muon statistics in about 5, 40, and 500 times, respectively. In every simulated event, information on energy depositions in scintillation detectors and on muon interactions with energy transfers more than 1 GeV was recorded. 

Analysis of simulation results has shown that qualitatively the selection parameters $E_{1},E_{2},K_{2}$ influence the event samples in a following way:
\begin{itemize}
\item the shift in $E_{2}$ is nearly proportional to the shift in muon energy;
\item increase of the minimal value of the relative depth of the deep $K_{2}$ suppresses contribution of nuclear showers (from inelastic muon interaction with nuclei) which may imitate multiple interactions;
\item increase of the threshold in $E_{1}$ decreases the number of muons with moderate energies ($\sim$TeV), while most of high-energy events (hundreds TeV) are retained.
\end{itemize}

Among possible versions of muon energy estimation in the pair meter technique, sufficiently effective and convenient is the use of rank statistics of energies transferred in muon interactions: transferred energies $\varepsilon_{j}$ in an individual event are arranged in a decreasing order, and $n$-th value $\varepsilon_{n}$ is then used to estimate muon energy \cite{6}. Energy depositions measured in scintillation planes of the telescope, which determine the longitudinal profile of the event, are not simply related with the transferred energies. This is caused by random location of interaction points relative to detector planes, superposition of cascades from different interactions, fluctuations of cascade development, etc. However, analysis of simulated events allows conclude that the energy deposition $E_{2}$ in the second in value maximum is determined mainly by the second in energy cascade, related with production of $e^{+}e^{-}$ pair by muon (relative energy transfers $\varepsilon/E_{\mu}\sim 10^{-2}-10^{-3}$), while the largest cascade (associated with the largest energy deposition $E_{1}$) with a high probability is caused by muon bremsstrahlung or inelastic muon interaction (with  $\varepsilon/E_{\mu}\geq 0.1$). Since the spectra of rank statistics are nearly similar to the spectrum of muons, it is expedient to use for the following analysis the distributions of events in the value of $E_{2}$, and to vary other parameters of event selection: $E_{1}$ ($\geq$ 5 GeV, $\geq$ 20 GeV, $\geq$ 40 GeV, etc.) and $K_{2}$ ($\geq$ 1, 2, 5, ...).

\section{Analysis of experimental data on multiple interactions of muons}

Experimental data accumulated at BUST during 12.5 years in 1983-1995 and 2 years (2003-2004) after restoration of amplitude measurement system \cite{13} have been analyzed. Periods of reliable operation of all systems responsible for energy deposition measurements were selected on the basis of a careful statistical analysis of the data. As a result, the total ``live'' time of registration amounted to $3.3\times 10^{8}$ s (more than 10 years), and the total number of events after preliminary selection (with total energy deposition $\geq$ 10 GeV in horizontal planes of the telescope) was about 10 millions. In more details, event selection criteria are described in \cite{14}. Only information of horizontal telescope planes was used. The total number of experimental events with twofold muon interactions selected with conditions $E_{1},E_{2}$ $\geq$ 5 GeV and muon tracks crossing all four horizontal planes equals to 1831; the corresponding statistics of simulated events amounts to 26951 events.

Experimental distributions of the events $N(E_{2})$ were compared with Geant4 simulation results for different selection criteria ($E_{1}\geq 5$ GeV and $K_{2}\geq 1$; $E_{1}\geq 20$ GeV and $K_{2}\geq 2$, etc.) and four different muon energy spectrum models (Fig. 3):
\begin{enumerate}
\item usual muon spectrum from $\pi$-, $\textit{K}$-decays in the atmosphere (equation (1) with $R=0$ and $\gamma_{\mu}=2.7$);
\item usual spectrum with addition of prompt muons at the level of $R=1\times10^{-3}$;
\item the same, but with three times higher prompt muon contribution, $R=3\times10^{-3}$;
\item usual spectrum with inclusion of VHE muons according to equation (2) with following parameters: \\ 
$n_{\mu}$ = 1, $f_{\mu}$ = 0.025, $E_{k}$ = 5 PeV, $\gamma_{1}=1.7$, and $\gamma_{2}=2.0$.
\end{enumerate}

\begin{figure}[h!]
\center \includegraphics[scale=0.85]{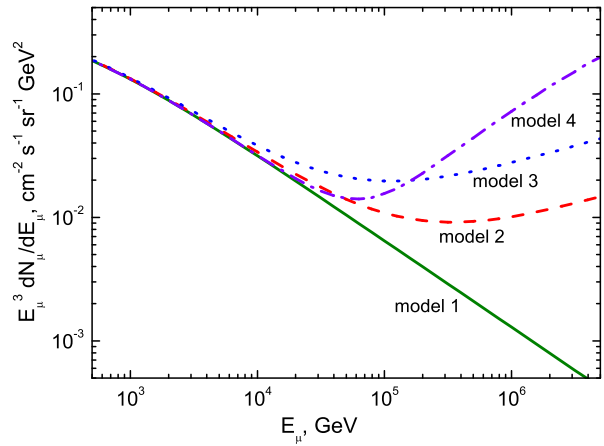}
\caption{Differential muon energy spectra for vertical direction (4 models).}
\end{figure} 

Experimental and calculated integral distributions of the events in $E_{2}$ are presented in Fig. 4. As a whole, within statistical uncertainties the data and calculations for a usual muon spectrum are in a good agreement in the range 5 GeV $\leq E_{2} \leq$ 30 GeV. However, at large values of $E_{2}$ (more than 80 GeV) the expected number of events is several times less (and at the tale of the distribution, almost ten times) than the observed in the experiment. Let us note that namely in the region $E_{2} \sim$ 100 GeV and higher the multiple interaction method in BUST becomes the most sensitive to the changes of muon spectrum shape.

\begin{figure}[h!]
\center
\includegraphics[scale=0.85]{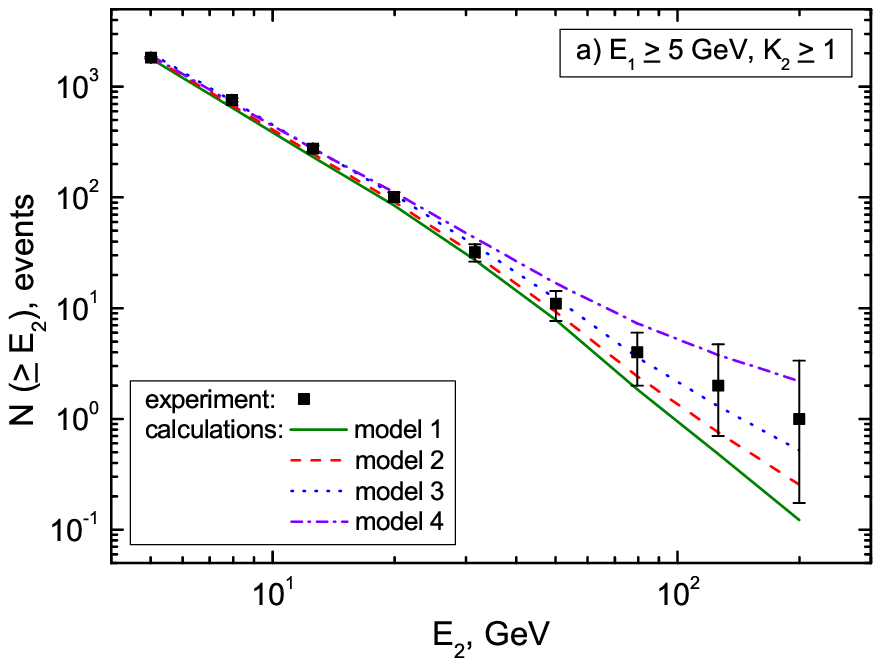}
\hspace{+10.0pt}
\includegraphics[scale=0.85]{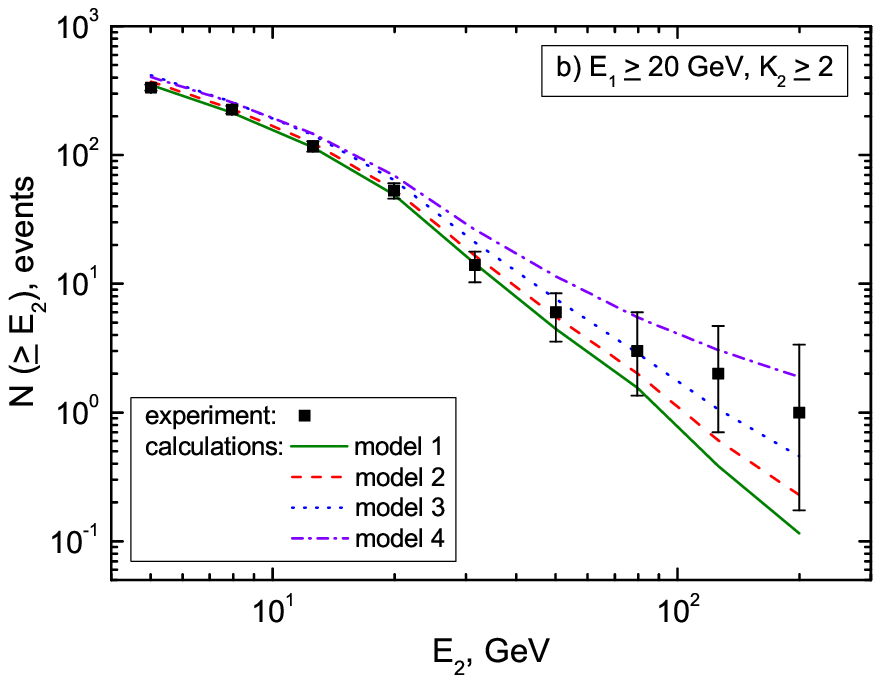}
\caption{Integral distributions of experimental events (the points) and expected spectra (the curves) in $E_{2}$ for 4 different muon spectrum models (see the text) and two sets of selection criteria (a,b).}
\end{figure} 

At comparison of the corresponding to Fig. 4a differential distribution in $E_{2}$ with the expected one under assumption of a usual muon spectrum (from $\pi$-, and $\textit{K}$-decays) the value of $\chi^{2}$ appears equal to 32.9 (at 8 degrees of freedom) which implies the rejection of such hypothesis on the spectrum shape with about $99.9\%$ confidence. Situation remains nearly the same after inclusion of prompt muons with $R=10^{-3}$ (spectrum model 2, $\chi^{2}=24.7$). Much better agreement is reached at comparison of the data with calculation results for sufficiently large fraction of prompt muons (model 3, $R=3\times10^{-3}$) or addition of VHE muons (model 4); corresponding values of $\chi^{2}$ in these cases are equal to 17.4 and 15.6, respectively.

It is important to note that the observed excess of events with large values of $E_{2}$ is retained at different approaches to data analysis and different selection criteria (compare Fig. 4a and Fig. 4b). Four events with highest values of $E_{2}$ (more than 80 GeV) are presented in Fig. 5. All these events are detected inside the telescope in all horizontal planes and have a clear topology. Therefore, in spite of low statistics, the deviation of experimental distributions from calculations performed in frame of generation of muons only in $\pi$-, $\textit{K}$-decays seems to be significant, and evidences for a possible existence of the fluxes of VHE or prompt muons with the considered parameters.

\begin{figure}[h!]
\center
\includegraphics[scale=0.5]{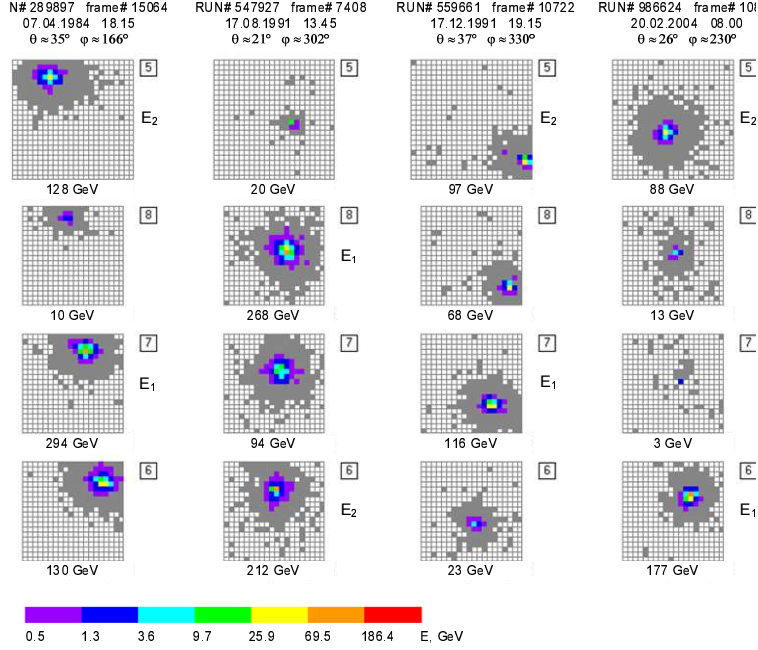}
\caption{Experimental events with highest values of $E_{2}$. Energy depositions measured in scintillation planes are indicated.}
\end{figure} 

\section{Muon energy spectrum}

In order to pass from the experimental distributions of event characteristics to the muon energy spectrum, it is necessary to determine which intervals of muon energies give the main contribution to generation of registered events, to choose effective estimates for them (mean, logarithmic mean, or median muon energies) and to define the conversion procedure.

Distributions of muon energies giving contribution to the events with several threshold values $E_{2}$ at fixed parameter $K_{2}$ calculated for 4 different assumptions on muon spectrum shape (spectrum models 1-4 described in the preceding section) are plotted in Fig. 6 (a,b,c,d). These distributions are rather wide even for a usual spectrum of muons (Fig. 6a), and, in presence of the additional muon flux with a more hard spectrum, at high $E_{2}$ values become bimodal (Figs. 6b-6d). The appearance of the second hump in the region of muon energies of hundreds TeV and higher is caused by a good sensitivity of the multiple interaction method namely to this, more hard, part of the muon spectrum.

\begin{figure}[h!]
\includegraphics[scale=0.875]{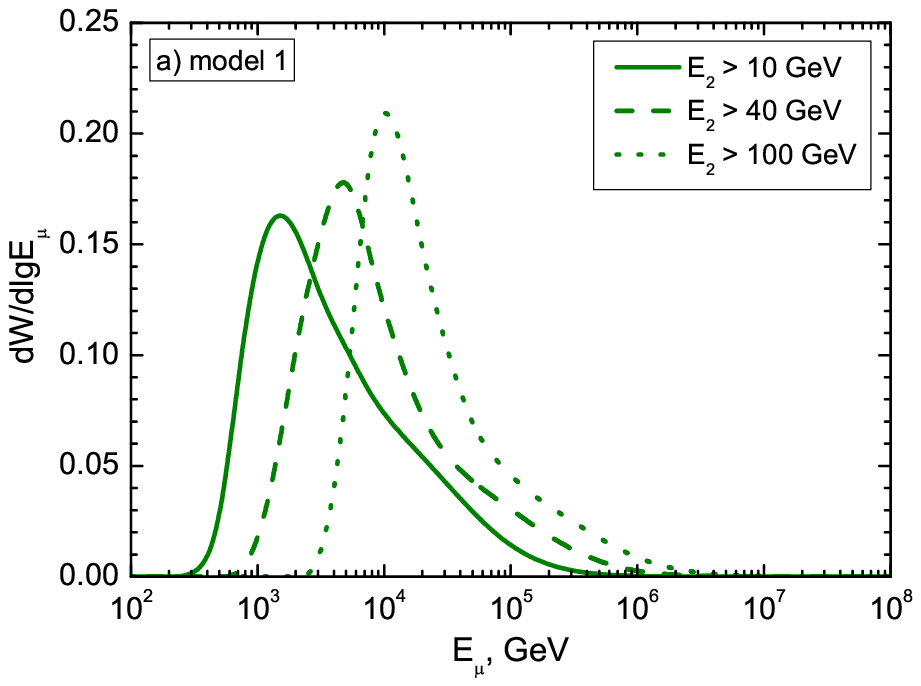}
\hspace{+1.0pc}
\includegraphics[scale=0.875]{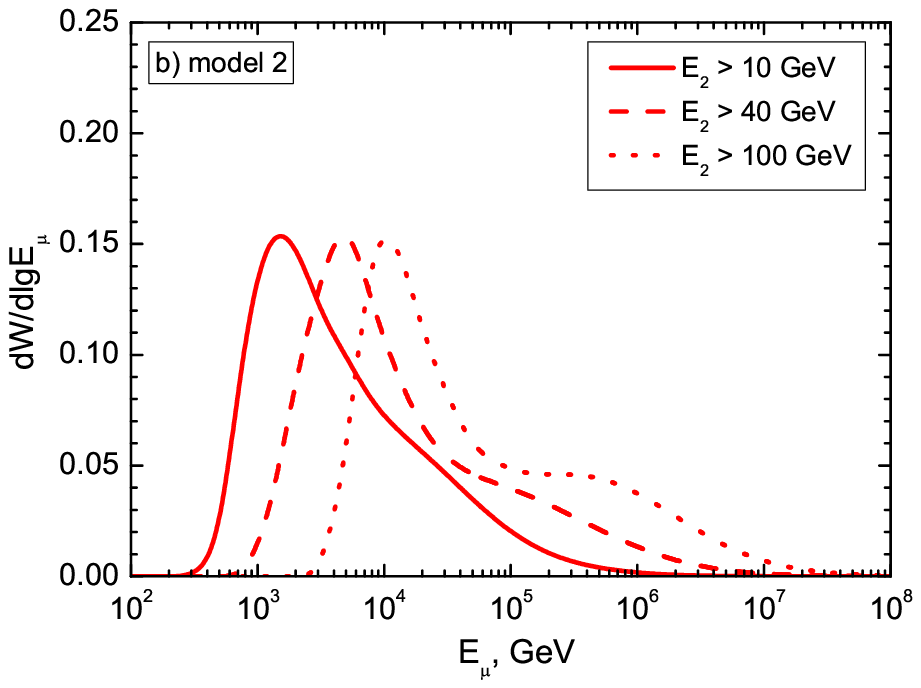}
\vspace{+1.0pc}
\includegraphics[scale=0.875]{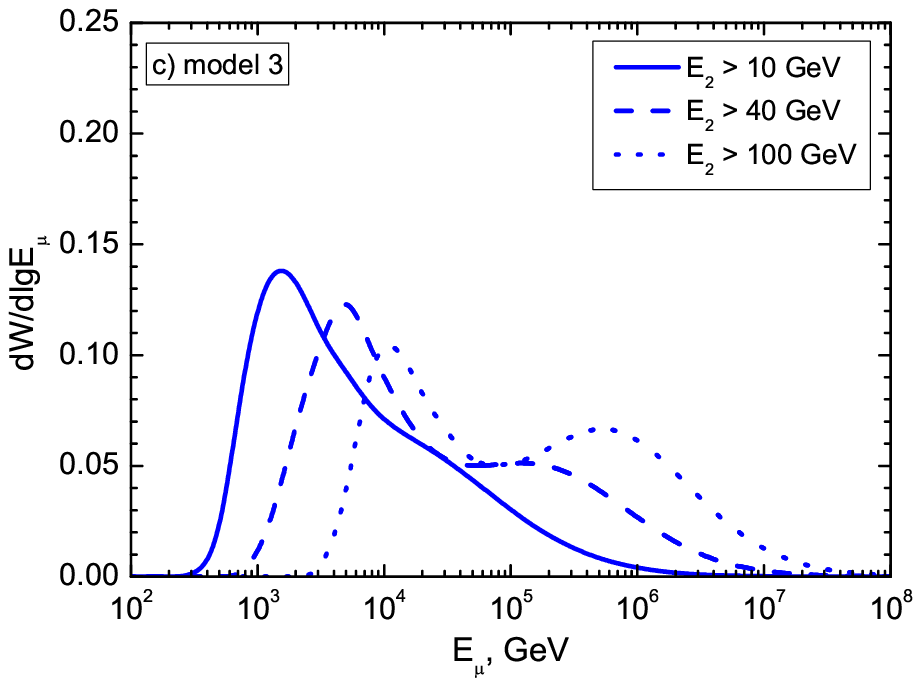}
\hspace{+1.0pc}
\includegraphics[scale=0.875]{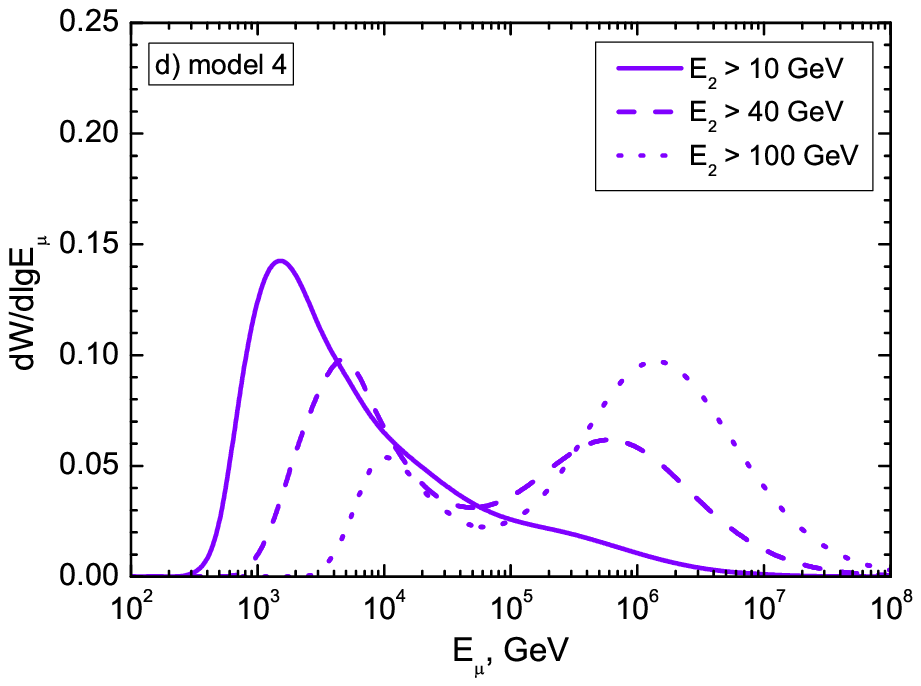}
\caption{Energy distributions of muons giving contribution to events with different threshold values $E_{2}$ for 4 models of muon energy spectrum.}
\end{figure}

In order to illustrate the decisive role of direct electron pair production process in the multiple interaction method, calculations for muon spectrum with the addition of VHE muons (model 4) were repeated with the exclusion of pair production. The obtained distributions (Fig. 7) appeared insensitive to the additional VHE muon flux (compare with Fig. 6d), and effective muon energies in this case would not exceed several tens TeV even for high values of $E_{2}$.

\begin{figure}[h!]
\center
\includegraphics[scale=0.875]{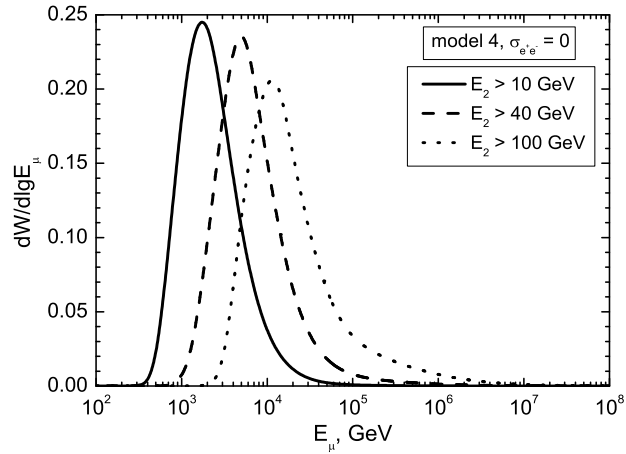}
\caption{Energy distributions of muons giving contribution to events with different threshold values $E_{2}$ for the model 4 of the muon spectrum with the ``switched-off'' pair production process (compare Fig. 6d).}
\end{figure} 

Energy spectra of muons from the BUST data on multiple interactions were obtained in a following way. At first, for certain sets of selection criteria ($E_{1}$ and $K_{2}$ parameters) the differential and integral distributions of the observed events $N_{\mathrm{obs}}$ with an equal step in common logarithm of $E_{2}$ were constructed, namely, the number of events in every bin $\Delta\lg$($E_{2}$, GeV) = 0.7-0.9, 0.9-1.1, ..., 2.3-2.5 for the differential distribution, and the total number of events with  $\mathrm{lg}$($E_{2}$, GeV) $\geq$ 0.7, $\geq$ 0.9, etc. for the integral one were counted. 

Expected model distributions $N_{\mathrm{mod}}$ of the events in $\mathrm{lg}E_{2}$, and also energy distributions of muons giving the contribution to events in a certain interval $\Delta\lg E_{2}$ or $\geq$ $\mathrm{lg}E_{2}$, corresponding mean, logarithmic mean, median energies $E_{\mu}^{*}$ for differential distributions and effective threshold muon energies $E_{\mu0}^{*}$ for integral ones were computed on the basis of the results of Geant4 telescope response simulations for the respective combination of selection criteria ($E_{1}$ and $K_{2}$) and four models of surface muon energy spectrum discussed above.

Dependences of logarithmic mean, mean, and median muon energies on $E_{2}$ (for differential in $E_{2}$ event distributions) are presented in Fig. 8 (a,b, and c respectively) for different spectrum models. These dependences clearly demonstrate the main advantage of the multiple interaction method, namely, the possibility to advance in muon energy region of hundreds TeV and even few PeV in case of the presence of substantial flux of muons with a hard spectrum at these energies. Since the energy deposition in scintillator layers of BUST constitutes about $10\%$ of cascade energy \cite{10}, the ratio between effective muon energy $E_{\mu}^{*}$ and $E_{2}$ reaches in this case the order of $10^{3}$. In Fig. 8d, the dependences of effective threshold energy $E_{\mu0}^{*}$ (estimated via logarithmic mean values) for the integral distributions in $E_{2}$ are shown. Qualitatively, it is seen that the dependences in Figs. 8a and 8d only weakly differ from each other.

\begin{figure}[h!]
\includegraphics[scale=0.875]{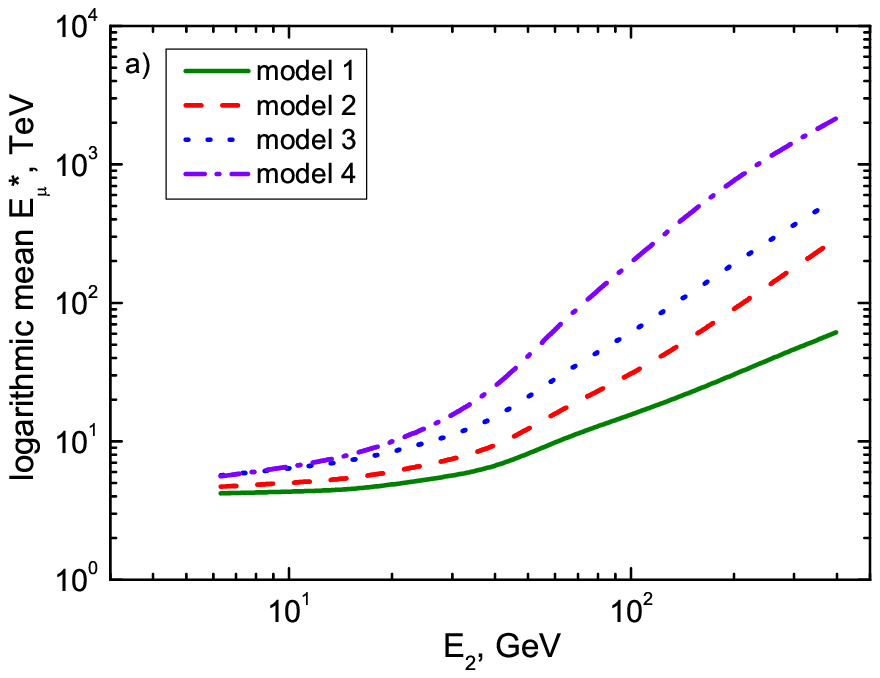}
\hspace{+10.0pt}
\includegraphics[scale=0.875]{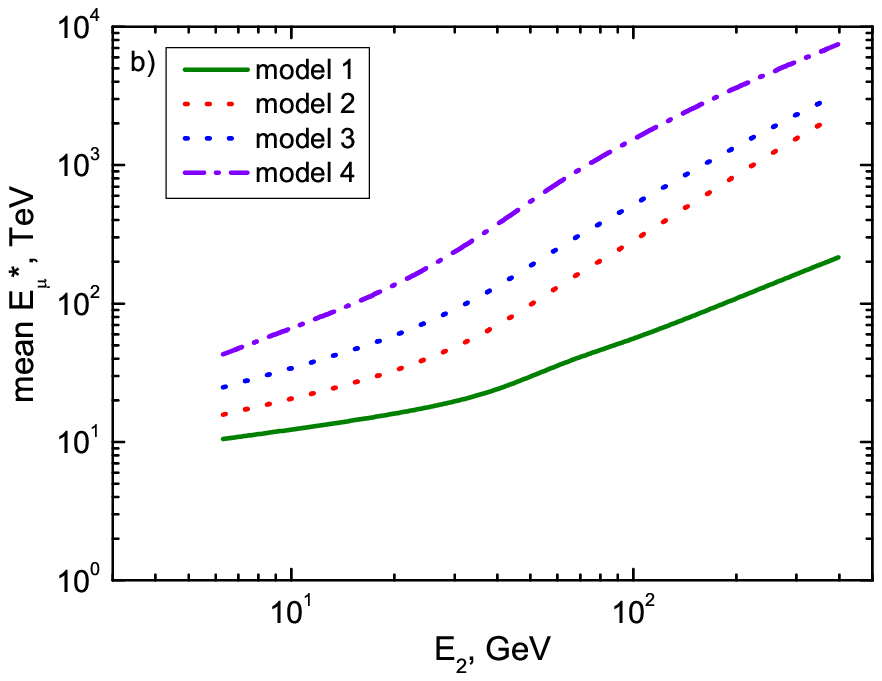}
\vspace{+1.0pc}
\includegraphics[scale=0.875]{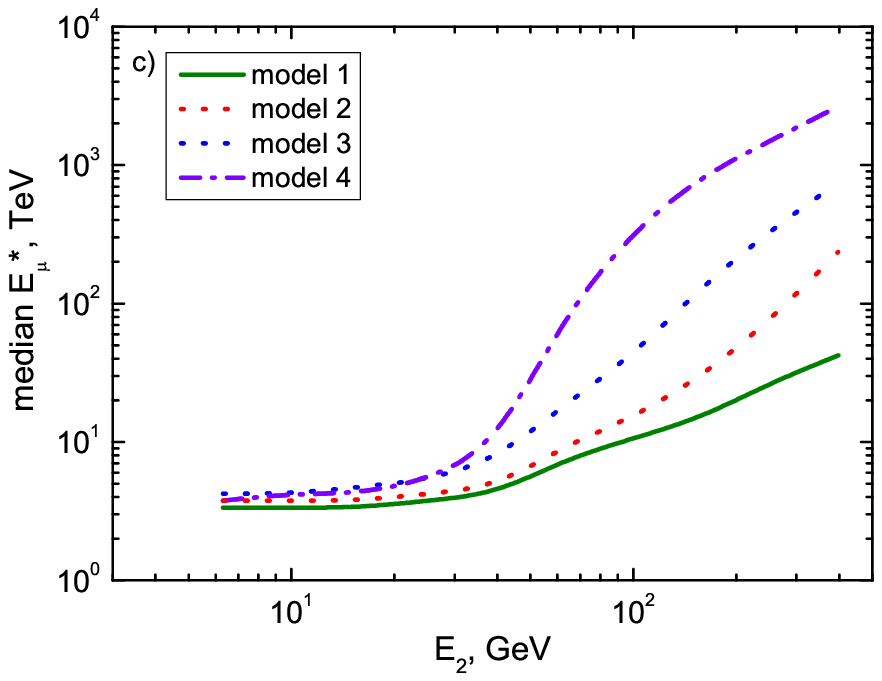}
\hspace{+20.0pt}
\includegraphics[scale=0.875]{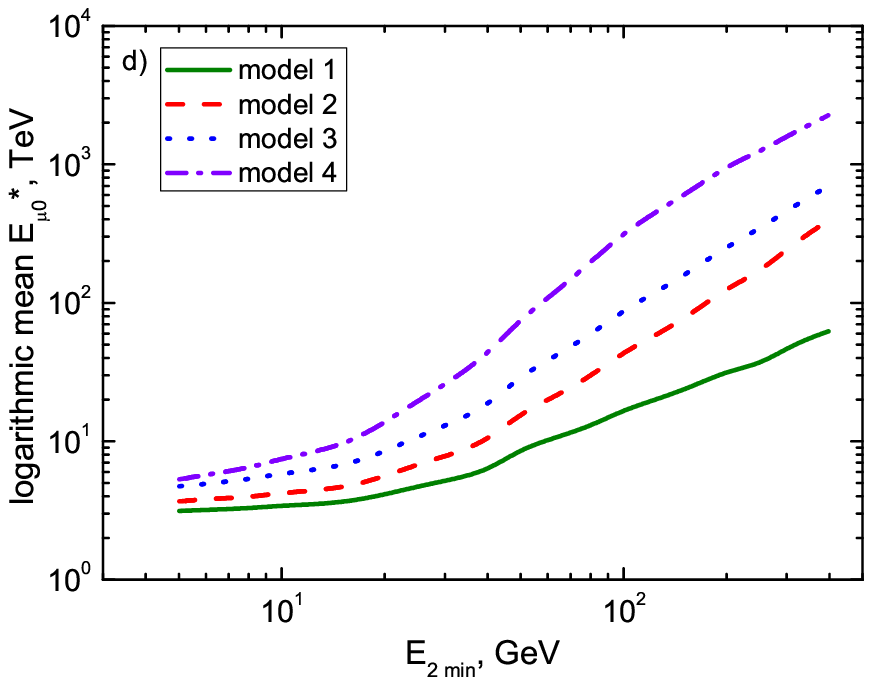}
\caption{Dependences of logarithmic mean (a), mean (b), and median (c) muon energies on $E_{2}$ for differential in $E_{2}$ distributions. Frame (d): effective threshold muon energies for integral in $E_{2}$ distribution. The curves correspond to four different spectrum models.}
\end{figure} 

Finally, the estimates of differential and integral muon spectra are found in a following way:

\begin{equation}
d\tilde{N}_{\mu}(E_{\mu}^{*})/dE_{\mu}=dN_{\mu}(E_{\mu}^{*})/dE_{\mu}\times N_{\mathrm{obs}}^{\textrm{dif}}(E_{2})/N_{\mathrm{mod}}^{\textrm{dif}}(E_{2}),
\end{equation}
\begin{equation}
\tilde{N}_{\mu}(\geq E_{\mu0}^{*})=N_{\mu}(\geq E_{\mu0}^{*})\times N_{\mathrm{obs}}^{\textrm{int}}(E_{2})/N_{\mathrm{mod}}^{\textrm{int}}(E_{2}),
\end{equation}
where $d{N}_{\mu}(E_{\mu}^{*})/dE_{\mu}$ and $N_{\mu}(\geq E_{\mu0}^{*})$ are differential and integral muon energy spectra for the respective spectrum model calculated at corresponding effective muon energy ($E_{\mu}^{*}$ and $\geq E_{\mu0}^{*}$).

Differential muon energy spectra for vertical direction reconstructed from the experimental data according to the described procedure at four different assumptions on muon spectrum model are presented in Fig. 9. Results are shown for one of the combinations of the selection criteria with highest statistics ($E_{1}\geq$ 5 GeV, $K_{2}\geq$ 1). Since there is no generally accepted definition of the effective energy of muons responsible for the observed events, the points corresponding to all three versions (mean, logarithmic mean, and median energies) are given in the figure. The curves in each frame represent the assumed spectrum models.

\section{Discussion}

The following conclusions can be made from the analysis of the results presented in Fig. 9. If one assumes that the muon spectrum is formed only due to decays of pions and kaons in the atmosphere (i.e. ``usual'' muon spectrum, Fig. 9a), then a strong dependence of spectrum reconstruction results on the choice of the effective muon energy (mean, logarithmic mean, median energy) appears as a large spread of reconstructed points. Furthermore, muon intensity estimated in frame of this assumption in the range of several tens TeV (considering median or logarithmic mean energy) or around 100 TeV (according to mean energy) is practically ten times higher than the expected one, and seriously contradicts results of other experiments compilation of which is given in [5].

\begin{figure}[h!]
\includegraphics[scale=0.9]{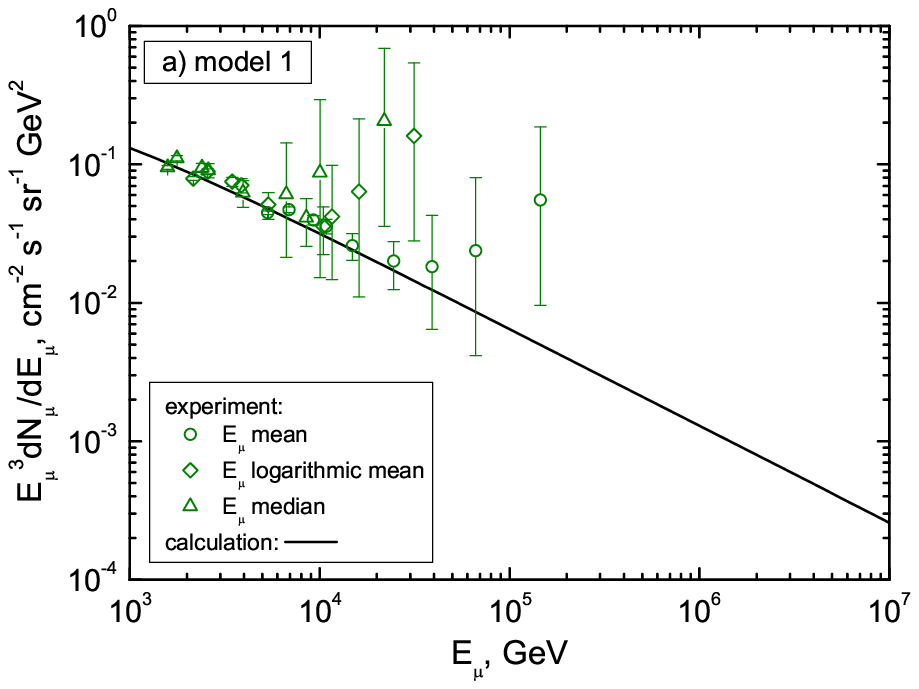}
\vspace{-1.5pt}
\vspace{+0.5pc}
\includegraphics[scale=0.9]{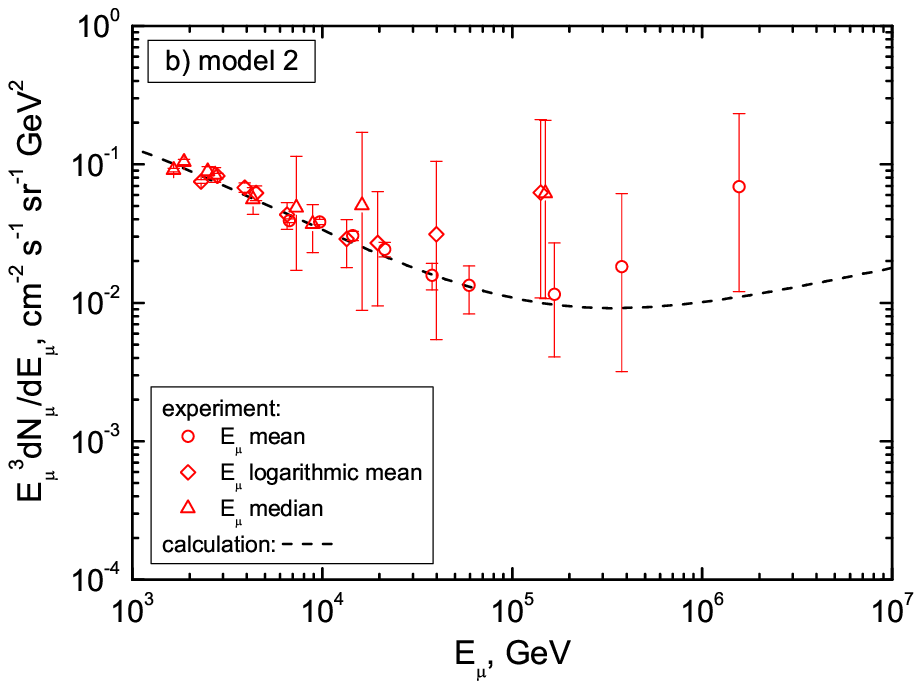}
\vspace{-1.5pt}
\vspace{+0.5pc}
\includegraphics[scale=0.9]{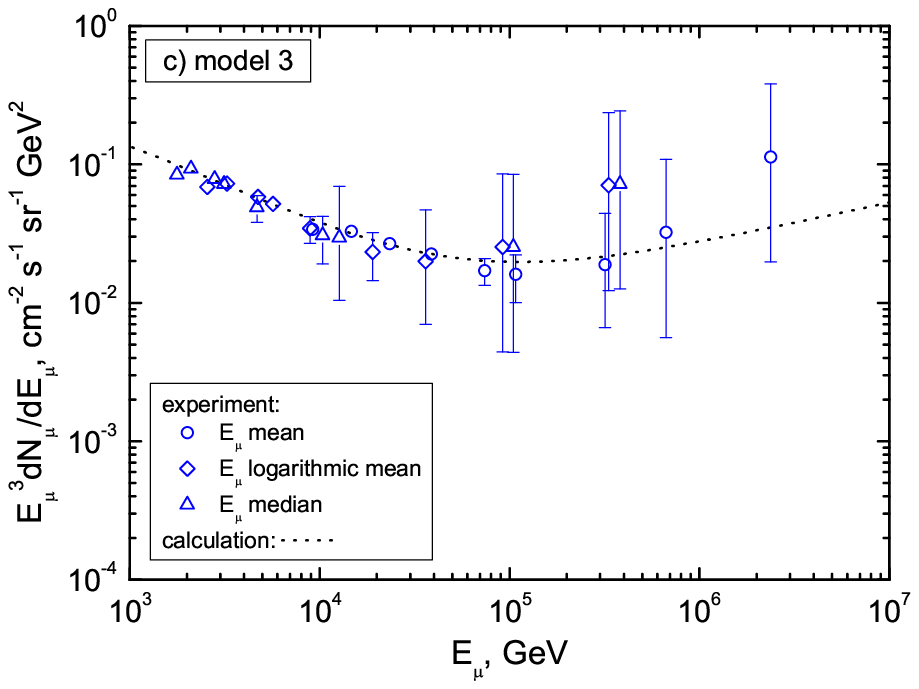}
\vspace{-1.5pt}
\vspace{+0.5pc}
\includegraphics[scale=0.9]{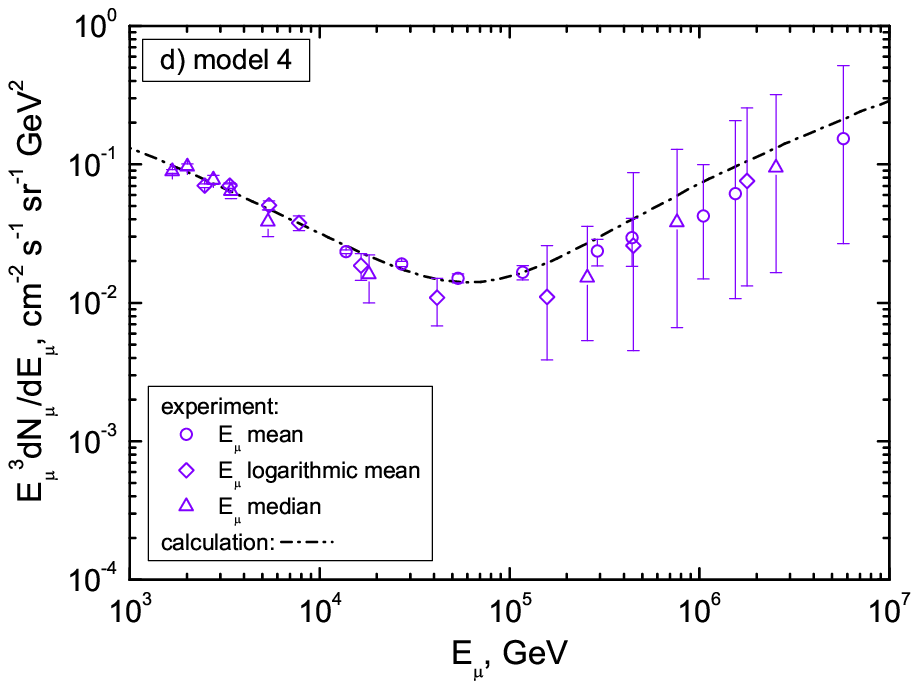}
\caption{Differential muon energy spectra reconstructed from BUST data on multiple interactions at different assumptions on muon spectrum model (a,b,c,d) with different choice of effective muon energy: mean (circles), logarithmic mean (diamonds), and median (triangles).}
\end{figure} 

The spread of experimental points relative to the model spectrum curves decreases as the contribution of additional muon flux with a more hard energy spectrum increases (Fig. 9b and 9c). At the same time, the agreement is improving also in the range of moderate muon energies (tens TeV); in other words, the dependence of results on the choice of effective muon energy (mean, logarithmic mean, median energy) at muon spectrum reconstruction disappears. The best agreement of the data with the expectation in a wide range of energies (from few TeV to few PeV) is observed for the spectrum with addition of the flux of VHE muons with parameters indicated above (Fig. 9d, spectrum model 4); the r.m.s. deviation of the points from the curve in this case is minimal.

In Fig. 10, the integral muon energy spectra measured at BUST by means of different methods are compared. One of the possible reasons of the difference between the results obtained from the depth-intensity curve and electromagnetic cascade spectrum measurements may be related with different procedures used for muon spectrum reconstruction from the experimental data. Thus, in the paper \cite{9}, in order to pass from the depth-intensity dependence (after evaluation of the $R$ parameter) to the integral energy spectrum of muons at the surface (taking into account muon energy loss fluctuations) the mean energy of that part of the spectrum which is responsible for muon flux intensity at a given depth was used. Authors note that the mean energy of prompt muons (from charmed particle decays), due to a more flat spectrum, is about twice more than the mean energy of usual muons; therefore weighted average of mean energies for two components of the flux was used for the conversion.

\begin{figure}[h!]
\center \includegraphics[scale=1.2]{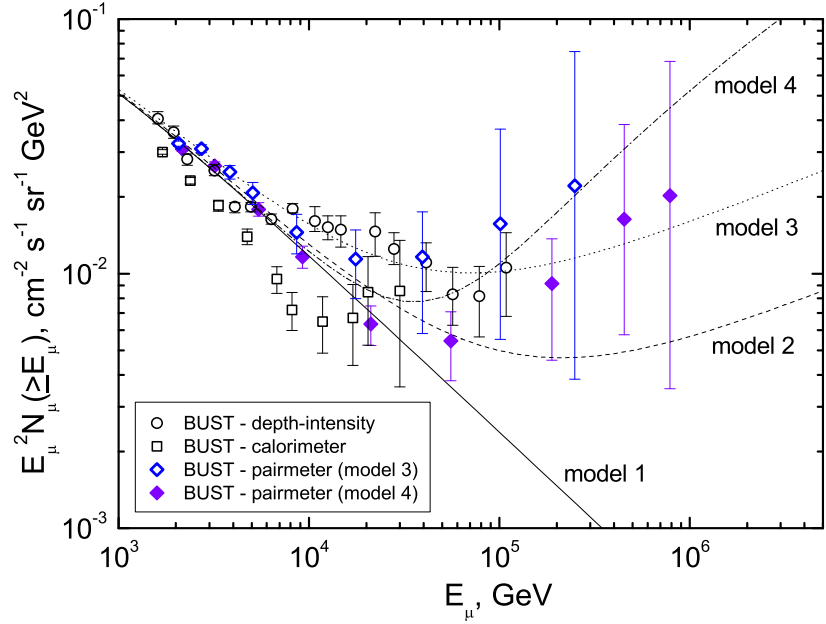}
\caption{Integral energy spectrum of muons for vertical direction reconstructed from the BUST data on depth-intensity curve (circles, \cite{9}), spectrum of electromagnetic cascades (triangles, \cite{10}), and by means of multiple interaction method for two models (3 and 4) used in muon spectrum reconstruction from experimental data. The curves represent calculations for different spectrum models.}
\end{figure} 

In the paper \cite{10}, the transition from the measured spectrum of energy depositions of electromagnetic cascades in the telescope to the muon energy spectrum was performed for median energies of muons responsible for events in a given energy deposition bin. A deep in the reconstructed muon energy spectrum around 10 TeV (Fig. 10) most probably is related with some methodical reasons, since it is difficult to suggest any physical explanation of its appearance. As to the absolute value of muon flux measured by this method, it is necessary to note that the systematic uncertainty in muon intensity could reach about $25\%$, since, as it was indicated in \cite{10}, the accuracy of the absolute energy calibration of energy deposition measurements was about $10\%$.

For the reconstruction of the integral muon energy spectrum from the data on multiple interactions of muons in BUST, the spectrum models 3 and 4 were used (open and solid diamonds in Fig. 10, respectively). The estimates of effective threshold muon energies were obtained on the basis of logarithmic mean values as optimal ones for quasi-power spectra of particles \cite{6}. As it is seen from the figure, no deviations from the usual spectrum is observed up to energies of at least $\sim$ 10 TeV for model 3 and $\sim$ 30 TeV for model 4, while around 100 TeV and higher a considerable excess in comparison with the spectrum of muons from $\pi$-, $\textit{K}$-decays appears. At that, the muon energy reconstruction by means of model 4 gives better agreement of experimental points with theoretical curve than by means of model 3.

A natural question may arise at discussions of the obtained results: how was it possible to register PeV muons with a relatively small-size setup ($\sim 200\:\mathrm{m^{2}})$ bearing in mind that their flux is extremely low? And these doubts are correct for usual decay muons; in this case, during the period of the experiment (more than 10 years), at best, one or two muons with energy above 1 PeV could cross the telescope. However, for more flat spectra (muons from charmed particles or VHE muons from new generation processes) the expected number of such muons may reach several tens. And, even taking into account relatively low probability of generation of events with twofold muon interactions in such a thin setup as BUST (of the order of $\sim 10^{-1}$), the possibility of registration of PeV muons becomes quite real.

In Fig. 11, the differential muon energy spectrum obtained from the BUST data by means of multiple interaction method is compared with results of other experiments taken from the compilation \cite{5}. As it is seen from the figure, the present data are the first ones at energies above 100 TeV and, in spite of low statistics, evidence for a change of muon spectrum behavior namely in this region. One may expect that this energy region will be accessible soon for investigations by means of cascade shower spectrum measurements at IceCube \cite{15}, and the use of pair meter technique at such scale setups would also allow to explore the range of PeV muon energies.

\begin{figure}[h!]
\begin{minipage}[h!]{1.0\linewidth}
\vspace{+1.5pc}
\center{\includegraphics[scale=0.55,angle=270]{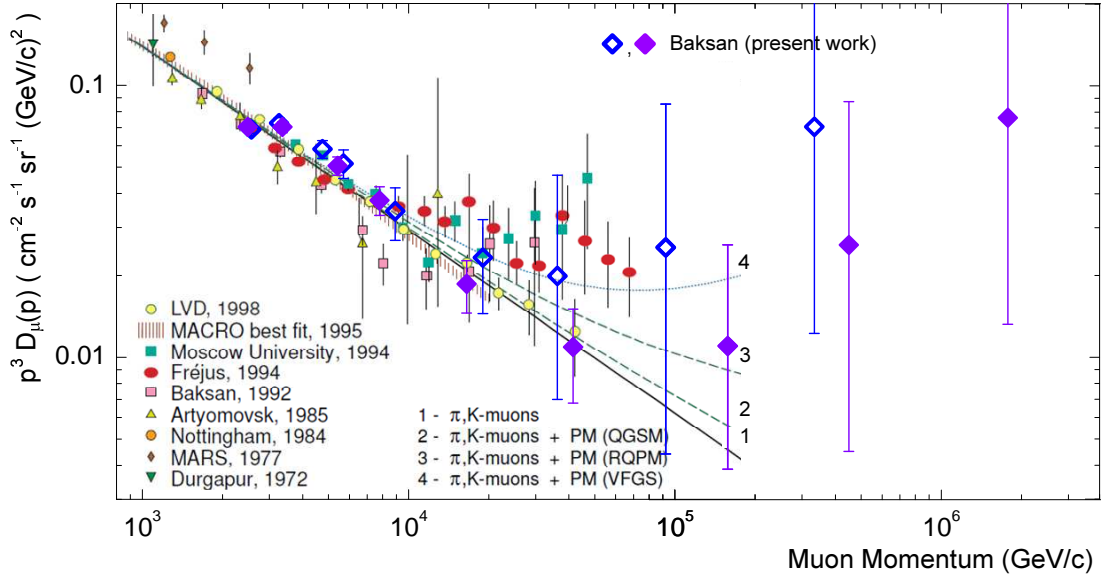}}
\end{minipage}
\caption{Differential muon energy spectra for vertical direction measured in various experiments (compilation from \cite{5}). The curves correspond to different models of prompt muon contribution considered in \cite{5}. Present BUST results obtained by means of multiple interaction method are added (open and solid diamonds for models 3 and 4 correspondingly).}
\end{figure} 

\section*{Conclusion}

Method of multiple interactions of muons based on the ideas of the pair meter technique gives possibility to use BUST data for estimation of the energy spectrum of cosmic ray muons in a wide energy region from several TeV to hundreds TeV. The analysis shows that no serious deviations from the usual spectrum formed as a result of pion and kaon decays are observed up to muon energies $\sim$ 20 TeV for model 3 and $\sim$ 50 TeV for model 4, if the existence of an additional flux of muons with a more hard spectrum is taken into account. At energies $\gtrsim$ 100 TeV this additional flux exceeds the expected contribution of muons from charmed particles corresponding to the parameter $R\sim 10^{-3}$, and may be explained with $R\sim 3\times10^{-3}$, which suggests a more fast increase of charmed particle yield compared to recent theoretical predictions. However, the best description of the experimental data can be achieved by assuming an additional contribution of VHE muons from new physical processes related with the appearance of the observed knee in cosmic ray energy spectrum.

\end{document}